\documentclass[%
 reprint,prf,
 amsmath,amssymb,
 aps,
]{revtex4-2}

\usepackage{graphicx}
\usepackage{dcolumn}
\usepackage{bm}
\usepackage{tikz}
\usetikzlibrary{arrows}
\usetikzlibrary{math}

\begin{document}


\title{Viscous Instabilities in Transversely Strained Channel Flows}

\author{Muhammad Abdullah}
 \email{muhabd@seas.upenn.edu}
\affiliation{Department of Mechanical Engineering, University of Pennsylvania
}%

\date{\today}

\begin{abstract}
We investigate here linear stability in a canonical three-dimensional boundary layer generated by the superposition of a spanwise pressure gradient upon an otherwise standard channel flow. As the main result, we introduce a simple coordinate transformation that enables the complete description of modal and non-modal stability using previous results on Poiseuille flow. We leverage this insight to derive closed forms for some relevant stability metrics. In particular, the critical Reynolds number for exponential-in-time growth is found to monotonically decrease with the strength of the cross-flow. A suitably chosen re-scaling, however, shows that the stability characteristics ultimately approach those of channel flow, despite the presence of a non-zero spanwise shear. Unstable eigenmodes akin to the Tollmien-Schlichting wave are found to propagate along the direction of the net flow. From a non-modal perspective, the maximal transient (algebraic) growth increases quadratically with the spanwise pressure differential and, similar to two-dimensional flows, is fueled by the lift-up effect. In this regard, the linear energy budget highlights a dramatic increase in energy production against the spanwise shear.
\end{abstract}

\maketitle

\section{\label{sec:introduction} Introduction}

Three-dimensional boundary layers (3DBLs) arise across numerous practical flow configurations, ranging from aerospace to geophysical, yet a robust description of their physics is far from complete. Most modern turbulence theories derive from statistically two-dimensional fluid motion and, therefore, offer limited insight on the three-dimensional counterpart, for which many foundational assumptions do not trivially carry over. Here, we investigate primary instabilities instigating laminar-turbulent transition in so-called \textit{transversely strained} channel (TSC) flows. Described by an auxiliary pressure differential across the span, these flows from prototypical test cases for theoretical and numerical investigations of wall-bounded 3DBLs; see, for example, \cite{moin_mansour, lozano_transition, lozano-durán_giometto_park_moin_2020}. Despite this -- and to the best of our knowledge -- laminar-turbulent transition in TSC flows has yet to be investigated, even in any preliminary sense, and this is where the bulk of the contribution of this work lies.

To begin, we reiterate that the turbulence in 3DBLs is fundamentally distinct from that in two-dimensional boundary layers. Canonically, 3DBLs are defined by skewed mean velocity vectors and, therefore, a \textit{cross-flow} persisting orthogonal to the local streamwise direction \cite{flack}. More specifically, the mean flow angle becomes a non-trivial function of the wall-normal coordinate, an anisotropy that enforces interesting and substantial departure from mean two-dimensional turbulence. For example, it is well-known that the production of turbulent kinetic energy in 3DBLs exhibits a significant decline relative to the two-dimensional case despite an increase in mean strain. In a similar vein, Townsend's structure parameter is also often observed to decrease, reinforcing the declining efficiency of the flow in producing turbulent energy \cite{moin_mansour}. Various proposals attempt to explain this through the cross-flow inhibiting large-scale vortical structures from generating Reynolds stresses \cite{bradshaw_pontikos_1985}. In contrast, \cite{lozano-durán_giometto_park_moin_2020} attribute this phenomenon, at least in TSC flows, to variations in pressure-velocity correlations, which drive a decrease in the production of normal stresses and, consequently, the turbulent energy.

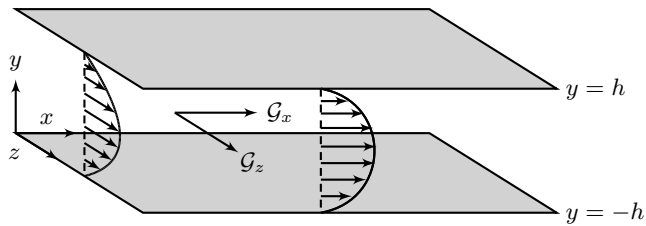
\begin{figure*}
    \centering
    \begin{tikzpicture}[scale = 1.1]

\coordinate (A) at (2, 2, 0.5);
\coordinate (B) at (3, 2, 1.65);
\coordinate (C) at (3, 2, 0.5);
    
\draw[thick, fill = gray!35] (0,0,0) -- (5,0,0) -- (7.5,0,2.5) node[right] {$y=-h$} -- (2.5,0,2.5) -- cycle;
\draw[thick, fill = gray!35] (0,1.5,0) -- (5,1.5,0) -- (7.5,1.5,2.5) node[right] {$y=h$} -- (2.5,1.5,2.5) -- cycle;

\draw[thick, -latex'] (0,0,0) -- node[midway, above] {$x$} (0.75,0,0);
\draw[thick, -latex'] (0,0,0) -- (0,0.65,0) node[above] {$y$};
\draw[thick, -latex'] (0,0,0) -- node[midway, left, xshift = -0.1cm, yshift = -0.1cm] {$z$} (0.85, 0, 0.85);

\draw[thick, -] (4.65,1.5,2.5) to[out = -10, in = 10, looseness = 1.5] (4.65, 0, 2.5) node[xshift = 1.85cm, yshift = 0.25cm, midway] {};

\draw[thick, -] (4.65,1.5,2.5) to[out = -10, in = 10, looseness = 1.5] (4.65, 0, 2.5);

\tikzmath{\N = 99;} 

\foreach \n in {0, 1, ..., \N}
{

\tikzmath{\y1 = 1.5 / (\N + 1) * \n;} 
\tikzmath{\y2 = 1.5 / (\N + 1) * (\n + 1);} 

\tikzmath{\z1 = -1.242236 * \y1 * \y1 + 1.8633540 * \y1 + 1.35;} 
\tikzmath{\z2 = -1.242236 * \y2 * \y2 + 1.8633540 * \y2 + 1.35;} 

\draw[thick] (\z1, \y1, \z1) -- (\z2, \y2, \z2);

}



\draw[thick, dashed] (4.65,1.5,2.5)--(4.65,0,2.5);
\draw[thick, dashed] (1.35,1.5,1.35) -- (1.35, 0, 1.35);

\draw[thick, -latex'] (2.5, 0.825, 1.5) -- (3.5, 0.825, 1.5) node[right] {$\mathcal{G}_x$};
\draw[thick, -latex'] (2.5, 0.825, 1.5) -- (3.75, 0.825, 2.75) node[right, xshift = -0.1cm, yshift = -0.135cm] {$\mathcal{G}_z$};

\draw[thick, -latex'] (4.65,0.2,2.5) -- (5.025,0.2,2.5);
\draw[thick, -latex'] (4.65,0.4,2.5) -- (5.2,0.4,2.5);
\draw[thick, -latex'] (4.65,0.6,2.5) -- (5.31,0.6,2.5);
\draw[thick, -latex'] (4.65,0.8,2.5) -- (5.325,0.8,2.5);
\draw[thick, -latex'] (4.65,1.025,2.5) -- (5.285,1.025,2.5);
\draw[thick, -latex'] (4.65,1.2,2.5) -- (5.175,1.2,2.5);
\draw[thick, -latex'] (4.65,1.35,2.5) -- (4.985,1.35,2.5);

\draw[thick, -latex'] (1.35,0.2,1.35) -- (1.67298,0.2,1.67298);
\draw[thick, -latex'] (1.35,0.4,1.35) -- (1.89658,0.4,1.89658);
\draw[thick, -latex'] (1.35,0.6,1.35) -- (2.02081,0.6,2.02081);
\draw[thick, -latex'] (1.35,0.8,1.35) -- (2.04565,0.8,2.04565);
\draw[thick, -latex'] (1.35,1,1.35) -- (1.97112,1,1.97112);
\draw[thick, -latex'] (1.35,1.2,1.35) -- (1.7972,1.2,1.7972);
\draw[thick, -latex'] (1.35,1.35,1.35) -- (1.60155,1.35,1.60155);


    \end{tikzpicture}
    \caption{A sketch of the present flow geometry. Respectively, $\mathcal{G}_x$ and $\mathcal{G}_z$ are the streamwise and cross-stream pressure gradients.}
    \label{fig:geometry}
\end{figure*}

Insofar as stability and transition is concerned, similarly dichotomous results can exist. Most recently, \cite{abdullah2024linear} examined modal and non-modal stability in the so-called oblique Couette-Poiseuille flows; they found that the introduction of three-dimensionality led to a strong decline in the potential for algebraic (transient) growth, although the critical Reynolds number for an exponentially-growing eigenmode decreased. Furthermore, the optimal configuration for short-time energy amplification was determined to be a collateral (i.e., essentially two-dimensional) boundary layer. For zero-pressure-gradient boundary layers superposed with a time-periodic spanwise wall motion, \cite{Hack_Zaki_2015} reported largely similar results on non-modal growth, attributing the declining amplification to modifications in the pressure-redistribution term in the linear energy budget. Previously, \cite{Blesbois_Chernyshenko_Touber_Leschziner_2013} had explored the effects of orthogonal wall motion on turbulent channel flow. They found that the ``generalized" optimal perturbation evolved via Orr amplification acting in tandem with the lift-up process. Meanwhile, \cite{corbett_bottaro_2001} found energy-optimal initial conditions in Falkner-Skan-Cooke boundary layers to develop exclusively through the lift-up effect. Moreover, they computed a stronger transient growth relative to the unswept case. Separately from those on rectilinear geometries, there exist various studies on flow in curvilinear coordinate systems, such as those on rotating cones, cylinders, and disks; important results on these configurations are summarized excellently in \cite{reed_saric_white}. 

Note that the addition of three-dimensionality usually allows for additional (and often unique) instability mechanisms outside of the standard Tollmien-Schlichting wave \cite{reed_saric_white}. For the present geometry, the turbulent-turbulent transition scenario was previously explored by \cite{lozano_transition}. These authors noted the development of a laminar-like spanwise boundary layer following the initial application of the transverse pressure gradient, whose strength subsequently dictated transition of this transient boundary layer due to either algebraic growth or the inviscid cross-flow instability.

We structure the remainder of this article as follows. Section \ref{sec:stability_problem} introduces the laminar state and the stability framework. Section \ref{sec:results} outlines an array of key results on both modal and non-modal stability. Finally, Section \ref{sec:conclusions} offers conclusions.

\section{\label{sec:stability_problem} The Stability Problem}

We begin by describing the background flow. The governing equations are the standard incompressible Navier-Stokes system, written in operator format as follows
\begin{equation}
\label{eqn:gov_eqs}
    \partial_t\bm{\xi} = \mathcal{F}\left(\bm{\xi}\right)
\end{equation}
where $\bm{\xi}$ is the state vector and $\mathcal{F}$ is a non-linear differential operator. A schematic describing the flow geometry is offered in Figure \ref{fig:geometry}. A standard channel flow driven by a pressure gradient $\mathcal{G}_x < 0$ is subject to an additional cross-stream pressure differential $\mathcal{G}_z < 0$, resulting in a so-called ``viscous-induced" three-dimensional boundary layer \cite{lozano-durán_giometto_park_moin_2020}. To scale the system, we adopt the channel half-height $h$ and the streamwise maximum $U_p$, respectively, as our characteristic length and velocity units. The adimensional laminar velocity profile is then given by
\begin{equation}
\label{eqn:base_state}
    \vec{U} = \begin{pmatrix}
        U\left(y\right) & 0 & W\left(y\right)
    \end{pmatrix}
\end{equation}
where
\begin{equation}
    U\left(y\right) = 1 - y^2 \qquad W\left(y\right) = \Pi\cdot U\left(y\right)
\end{equation}
and $\Pi\equiv \mathcal{G}_z/\mathcal{G}_x \geq 0$. The dimensionless Reynolds number $Re$ is defined as $Re = U_p h/\nu$, where $\nu$ is the kinematic viscosity.

We now linearize Equation (\ref{eqn:gov_eqs}) around the base state in the usual manner; see, for example, \cite{schmidstability}. In particular, by denoting $\vec{x} = \begin{pmatrix}
    x & y & z
\end{pmatrix}$, the state vector is hereafter decomposed as
\begin{equation}
    \bm{\xi}\left(\vec{x}, t\right) = \overline{\bm{\xi}}\left(\vec{x}\right) + \bm{\xi}^{\prime}\left(\vec{x}, t\right)
\end{equation}
where $\bm{\xi}^{\prime}$ represents a set of infinitesimal fluctuations superposed on the steady reference state $\overline{\bm{\xi}}$, described by Equation (\ref{eqn:base_state}). Then, a Jacobian linearization is performed around $\overline{\bm{\xi}}$, which yields the celebrated \textit{Orr-Sommerfeld-Squire} system. When recast in the traditional formulation involving the wall-normal velocity/vorticity formulation -- $v$ and $\eta$ respectively -- it reads

\begin{widetext}
    \begin{align}
\label{eqn:os}
    \left[\left(\partial_t + U\partial_x + W\partial_z\right)\nabla^2 - (\partial_y^2 U)\partial_x- (\partial_y^2 W)\partial_z- \dfrac{1}{Re}\nabla^4\right]v & = 0,\\
    \label{eqn:sq}
       \left[\partial_t + U\partial_x + W\partial_z - \dfrac{1} {Re}\nabla^2\right]\eta-(\partial_y W)\partial_x v+(\partial_y U)\partial_z v&= 0,
\end{align}
\end{widetext}
where $\nabla^2 = \partial_x^2 + \partial_y^2 + \partial_z^2$ and $\nabla^4$ is the bi-harmonic operator. Since the background flow is both streamwise and spanwise-homogeneous, we may assume a traveling-wave ansatz of the following form
\begin{equation}
    \begin{pmatrix}
    v\left(x,y,z,t\right) & \eta\left(x,y,z,t\right)
\end{pmatrix} = \begin{pmatrix}
    \hat{v}\left(y, t\right) & \hat{\eta}\left(y, t\right)
\end{pmatrix}e^{i\left(\alpha x+\beta z\right)}
\end{equation}
where $\alpha,\beta\in\mathbb{R}$ denote a set of spatial wavenumbers. Equations (\ref{eqn:os}) and (\ref{eqn:sq}) then become
\begin{equation}
\label{eqn:oss_ivp}
    \mathsf{L}\vec{q} = -\partial_t\mathsf{M}\vec{q},
\end{equation}
where $\vec{q} = \begin{pmatrix}
    \hat{v} & \hat{\eta}
\end{pmatrix}$ and by denoting $k^2 = \alpha^2+\beta^2$, we have defined
\begin{equation}\label{eqn:orr_somm}
    \mathsf{L} = \begin{pmatrix}
        \mathsf{L}_{\mathrm{OS}} & 0 \\ i\beta \partial_yU - i\alpha\partial_yW  & \mathsf{L}_{\mathrm{SQ}}
    \end{pmatrix},\qquad \mathsf{M} = \begin{pmatrix}
        \partial_y^2 - k^2 & 0 \\ 0 & 1
    \end{pmatrix}.
\end{equation}
The Orr-Sommerfeld and Squire operators, $\mathsf{L}_{\mathrm{OS}}$ and $\mathsf{L}_{\mathrm{SQ}}$, respectively, are given by

\begin{widetext}
    \begin{align}
\label{eqn:os_operator}
    \mathsf{L}_{\mathrm{OS}} & = \left(i\alpha U+i\beta W\right)(\partial_y^2-k^2) - i\alpha \partial_y^2 U-i\beta \partial_y^2 W-\dfrac{1}{Re}(\partial_y^2-k^2)^2, \\
    \label{eqn:sq_operator}
    \mathsf{L}_{\mathrm{SQ}} & = i\alpha U +i\beta W -\dfrac{1}{ Re}(\partial_y^2-k^2).
\end{align}
\end{widetext}

\begin{figure*}
    \centering
    \includegraphics[width=0.95\textwidth]{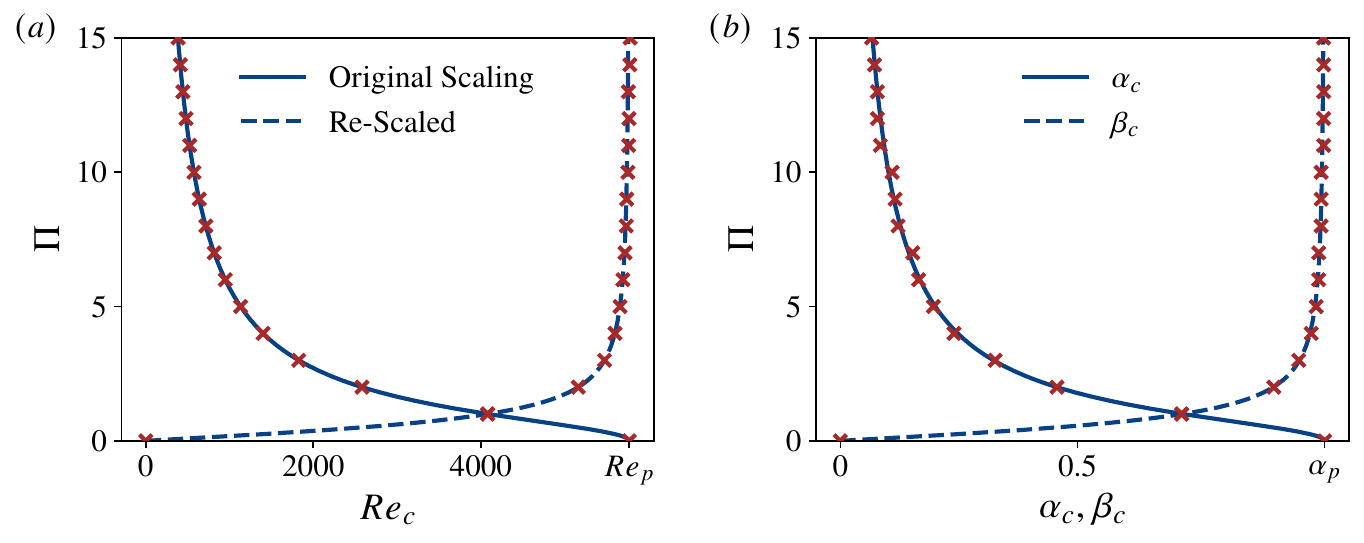}
    \caption{The critical parameters for eigenvalue instability in TSC flows plotted against $\Pi$; $(a)$, the critical Reynolds number -- the dashed line represents the re-scaled value involving the center-line velocity for the spanwise profile; $(b)$, the critical streamwise and spanwise wavenumbers. In both plots, the markers denote the critical parameters obtained via numerical experiments.}
    \label{fig:crit_params}
\end{figure*}
To ensure asymptotic stability, we require $\mathfrak{R}\left(\lambda\right) < 0$, where $\lambda\in\Lambda\left(\mathsf{S}\right)$, the eigenspectrum of $\mathsf{S}\equiv-\mathsf{M}^{-1}\mathsf{L}$. Traditionally, this condition is rephrased by setting $\lambda = -i\omega$, where $\omega\equiv \omega_r + i\omega_i$ is a circular frequency, so that the constraint for eigenvalue stability becomes $\omega_i<0$. However, since $\mathsf{S}$ has, in general, oblique eigenfunctions, large linear amplification -- often superseding modal growth -- can still occur over finite time horizons due to non-modal disturbances \citep{tref_pseudospec,schmidstability}. So as to capture this, we first introduce the energy norm (see, e.g., \cite{butler_3d_opt_pert})
\begin{equation}
    \left\lVert \vec{q}\,\right\rVert_E = \int_{-1}^1 \hat{v}^\dagger\hat{v} + \dfrac{1}{k^2}(\hat{\eta}^\dagger \hat{\eta} + \partial_y \hat{v}^{\dagger}\partial_y \hat{v})\,\mathrm{d}y
\end{equation}
and then consider the state transition operator $\Phi\left(t^{\prime}, 0\right)$ for Equation (\ref{eqn:oss_ivp}), which maps the initial state of the system at $t=0$ to its final state at $t = t^{\prime}$. Since $\mathsf{S}$ is time-invariant, the associated Peano-Baker expansion can be trivially summed to obtain
\begin{equation}
    \Phi\left(t,0\right) = e^{t\mathsf{S}}
\end{equation}
The operator norm $G\left(\alpha, \beta, Re, \Pi, t\right) \equiv \left\lVert \Phi\left(t, 0\right)\right\rVert_E^2$ then optimizes (at time $t$) the energy amplification incurred over all possible unit norm initial conditions, providing a measure for the pervasiveness of non-normality in the linear dynamics of the system. Via a similarity transformation, one can compute $G$ through a Schmidt decomposition (or in the discrete case, the more familiar singular value decomposition) by converting to a weighted 2-norm -- see \cite{TrefethenEmbree}. For numerical discretization of the Orr-Sommerfeld-Squire system, we employ the usual pseudo-spectral method based on Chebyshev polynomials \cite{tref_spec_methods}.

Before proceeding, we introduce a modified coordinate basis $\vec{x}^* = \begin{pmatrix}
    x^* & y^* & z^*
\end{pmatrix}$, where
\begin{equation}
\label{eqn:transformation}
    \begin{pmatrix}
    x^* \\ y^* \\ z^*
    \end{pmatrix} = \begin{pmatrix}
        \phantom{-}x\cos\phi + z\sin \phi \\ y \\ -x\sin\phi + z\cos\phi
    \end{pmatrix}
\end{equation}
and $\phi = \arctan\left(\Pi\right)$. Equation (\ref{eqn:transformation}) informs the following directional cosine operator
\begin{equation}
\label{eqn:dco}
    \mathsf{R} = \begin{pmatrix}
        \phantom{-}\cos\phi & 0 & \sin\phi \\ \phantom{-}0 & 1 & 0 \\ -\sin\phi & 0 & \cos\phi
    \end{pmatrix}
\end{equation}
from which it becomes apparent that $\vec{x}$ and $\vec{x}^*$ differ merely by a $\phi$-rotation (in a sense that maintains right-handedness) around the $y$-axis. The utility of this transformation is not immediately clear but follows from noting that despite $\vec{U}$ prescribing a three-dimensional boundary layer, the flow angle $\phi$, in fact, remains \textit{constant} in $y$; in particular, we have
\begin{equation}
    \tan\phi = \dfrac{W\left(y\right)}{U\left(y\right)} = \Pi \implies \phi = \arctan\left(\Pi\right)
\end{equation}
More importantly, within $\vec{x}^*$, the velocity profile $\vec{U}^*$ becomes uni-directional
\begin{equation}
\label{eqn:modified_profiles}
    \vec{U}^* = \begin{pmatrix}
        U^* & 0 & 0
    \end{pmatrix}
\end{equation}
where 
\begin{equation}
    U^*\left(y\right) = \sqrt{U\left(y\right)^2 + W\left(y\right)^2} = \Pi^*(1-y^2)
\end{equation}
and $\Pi^* = \sqrt{1 + \Pi^2}$. Furthermore, the wavenumber vector $\vec{k} = \begin{pmatrix}
    \alpha & \beta
\end{pmatrix}$ transforms as follows
\begin{equation}
\label{eqn:wavenumber_vector}
    \vec{k}^* \equiv \begin{pmatrix}
        \alpha^* \\ \beta^*
    \end{pmatrix} =
        \begin{pmatrix}
            \alpha\cos\phi + \beta\sin\phi \\ \beta\cos\phi - \alpha\sin\phi
        \end{pmatrix}
\end{equation}
This greatly simplifies the ensuing analysis.

\section{\label{sec:results} Results}

\subsection{\label{ssec:eigval_stability} Eigenvalue Stability}

To begin with, we define the critical Reynolds number $Re_c$ as the minimum Reynolds number below which the flow remains linearly stable to all infinitesimal disturbances. The associated critical wavenumbers are denoted by $\left(\alpha_c,\beta_c\right)$ and must achieve neutral stability. For background flows with a non-trivial spanwise component, it is well-known that Squire's Theorem is no longer applicable. In particular, although a ``two-dimensional'' eigenproblem can be constructed, no immediate information is obtained on the appropriate search space for instability \cite{mack_BL_stability}. Consequently, to fully characterize the stability portrait for TSC flows, a sweep is required throughout the entire $\left\{\alpha,\beta, Re,\Pi\right\}$-space. However, intuitively, one expects the stability features to approach those of the classic Poiseuille flow as $\Pi\to\infty$ (indeed, as $\Pi\to 0$); therefore, we are particularly interested in intermediate regimes.

Here, we make substantial use of Equation (\ref{eqn:transformation}). First, we note that Squire modes are necessarily damped, so that only the Orr-Sommerfeld operator is of relevance. When $\beta = 0$, we immediately recover the stability equations for spanwise-independent disturbances developing in a simple Poiseuille flow. More generally, take $\mathbb{E}\left(\omega, \vec{q}; \alpha, \beta, Re, U, W\right)$ to be the eigenproblem associated with the operator $\mathsf{S}\left(\alpha, \beta, Re, U, W\right)$. Using Equations (\ref{eqn:modified_profiles}) and (\ref{eqn:wavenumber_vector}), it is easy to recover the following relationship
\begin{equation}
\mathbb{E}\left(\omega, \vec{q}; \alpha, \beta, Re, U, W\right) = \mathbb{E}\left(\omega, \vec{q}; \alpha^*, \beta^*, Re, U^*, 0\right)
\end{equation}
With some additional effort, one can further find that
\begin{equation}
\label{eqn:reduced_evp}
\mathbb{E}\left(\omega, \vec{q}; \alpha^*, \beta^*, Re, U^*, 0\right) = \mathbb{E}\left(\omega^*, \vec{q}; \alpha^*, \beta^*, Re^*, U, 0\right)
\end{equation}
where $\omega^* = \omega/\Pi^*$ and $Re^* = \Pi^* Re$, so that all stability results from Poiseuille flow are now (indirectly) applicable. In particular, observe that Equation (\ref{eqn:reduced_evp}) is amenable to Squire's Theorem when applied to $\vec{x}^*$. Consequently, upon setting $\beta^* = 0$ and recalling that Poiseuille flow first becomes linearly unstable beyond $\left(\alpha_c^*, Re_c^*\right) = \left(\alpha_p, Re_p\right) \approx \left(1.02, 5772.72\right)$, we conclude that the critical Reynolds number for TSC flows satisfies the following closed form
\begin{equation}
    Re_c = \dfrac{Re_p}{\Pi^*} = \dfrac{Re_p}{\sqrt{1+\Pi^2}}
\end{equation}
Additionally, the critical wavenumbers (in the original coordinate frame) scale as
\begin{equation}
\label{eqn:critical_wavenumbers}
    \vec{k}_c = \begin{pmatrix}
        \alpha_c \\ \beta_c
    \end{pmatrix} = \dfrac{1}{\Pi^*}\begin{pmatrix}
        \alpha_p \\ \Pi\alpha_p
    \end{pmatrix} = \dfrac{1}{\sqrt{1+\Pi^2}}\begin{pmatrix}
        \alpha_p \\ \Pi\alpha_p
    \end{pmatrix}
\end{equation}
\begin{figure*}
    \centering
    \includegraphics[width=\textwidth]{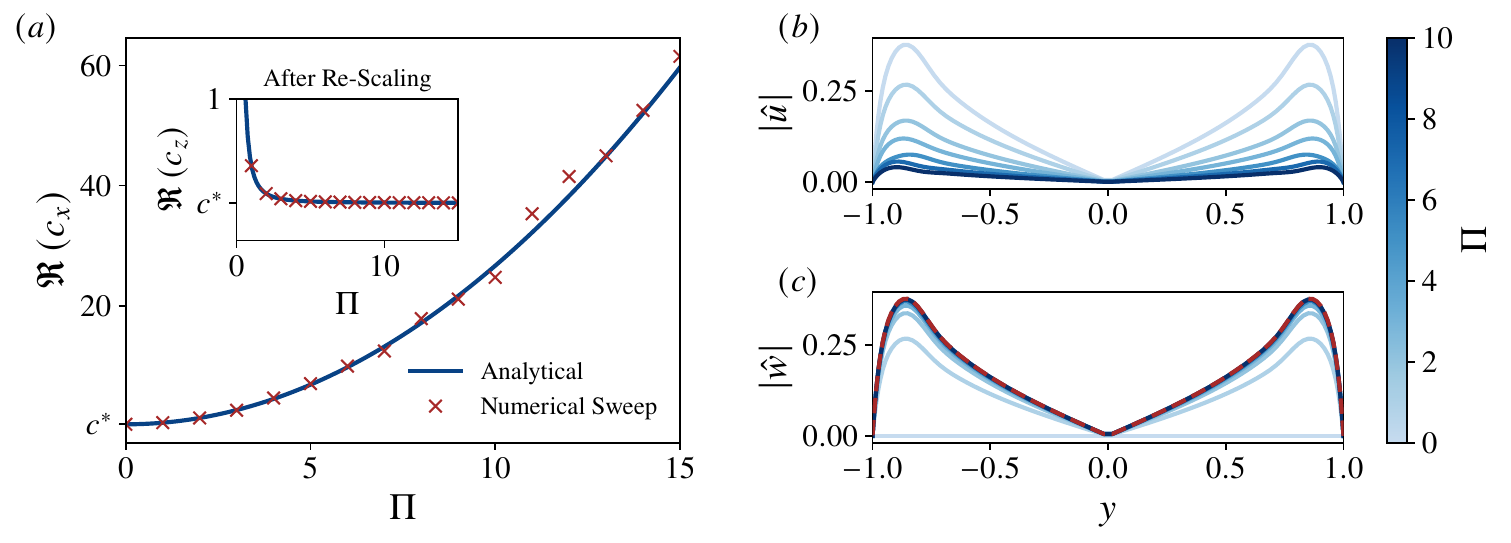}
    \caption{$(a)$, the critical (streamwise) phase speed plotted against $\Pi$; the inset shows the re-scaled spanwise phase speed $c_z$. Note that formally, both are real. Once again, the markers denote values obtained from a numerical sweep. Panels $(b)$ and $(c)$, respectively, depict amplitudes of the streamwise and spanwise components corresponding to the unstable eigenmode for TSC flows. In $(c)$, the dashed line denotes $u\left(\Pi\to 0\right)$.}
    \label{fig:crit_crs_and_efuncs}
\end{figure*}Some interesting remarks can now be made. Referencing Figure \ref{fig:crit_params}, we first note that $Re_c\to 0$ as $\Pi\to\infty$. At first glance, this appears erroneous but is merely a consequence of our non-dimensionalization. Specifically, if instead the characteristic velocity is chosen as $U_p\to \Pi U_p$ so as to coincide with the \textit{spanwise} maximum, we find $Re_c\to Re_p$ as $\Pi\to\infty$, which is in line with our earlier intuition. Physically, this limit corresponds to $U\approx0$ and $W\approx 1-y^2$, i.e. a simple parabolic flow along the span. Indeed, this is why $\beta_c\to \alpha_p$ in the same limit, since for large $\Pi$, the spanwise wavenumber begins to behave as a ``proxy" for the streamwise wavenumber.

Recall now that in Equation (\ref{eqn:reduced_evp}), the eigenvalue satisfies $\omega = \omega^*\Pi^*$. When written in terms of the phase speed $c_x$, we have $c_x = \omega/\alpha = \omega^*\Pi^*/\alpha$. Then, at criticality, using Equation (\ref{eqn:critical_wavenumbers}), we have
\begin{equation}
    c_x = \omega^*{\Pi^*}^2/\alpha_p = \omega^*(1+\Pi^2)/\alpha_p = c^*(1+\Pi^2)
\end{equation}
where $\mathfrak{I}\left(c\right) = 0$ (due to neutral stability) and $c^* \approx 0.264$ \cite{Criminale_Jackson_Joslin_2003}. Clearly, $c_x\to \infty$ as $\Pi\to\infty$, which follows from the fact that $\alpha\to 0$ in the same limit. In fact, from our observations earlier, a better metric here is the \textit{spanwise} phase speed, $c_z = \omega/\beta$. When coupled with the re-scaling $U_p\to\Pi U_p$ of our reference velocity, one can find
\begin{equation}
    c_z = \dfrac{c^*(1+\Pi^2)}{\Pi^2}
\end{equation}
so that
\begin{equation}
    \lim_{\text{$\Pi\to\infty$, Re-Scaled}}c_z = c^* = \lim_{\text{$\Pi\to0$}}c_x
\end{equation}
which, again, matches our intuition. For reference, these results have been plotted in Figure \ref{fig:crit_crs_and_efuncs}$(a)$.

Finally, from the transformed problem Equation (\ref{eqn:reduced_evp}), it is easy to see that an unstable eigenmode, if any, must correspond to the classic Tollmien-Schlichting (TS) wave for Poiseuille flow, traveling, however, in the direction of the \textit{net flow}. This disturbance has a vanishing spanwise component and, as such, can be written as $\begin{pmatrix}
    u^* & v^* & 0
\end{pmatrix}$ in the modified reference frame. Through Equation (\ref{eqn:dco}), we can then compute
\begin{equation}
    \begin{pmatrix}
        u \\ v \\ w
    \end{pmatrix} = \mathsf{R}^{-1}\begin{pmatrix}
    u^* \\ v^* \\ 0
\end{pmatrix} = \dfrac{1}{\Pi^*}\begin{pmatrix}
    u^* \\ v^* \\ \Pi u^*
\end{pmatrix}
\end{equation}
Thus, in the limit that $\Pi\to\infty$, $u\to 0$ and $w$ approaches the streamwise component of the TS perturbation, i.e. $w\left(\Pi \to \infty\right) = u\left(\Pi\to 0\right)$. This is verified in Figure \ref{fig:crit_crs_and_efuncs}$(b)$ and provides additional confirmation that the stability characteristics of TSC flows smoothly vary to and from those of Poiseuille flow.

\subsection{\label{ssec:transient_growth} Transient Growth}

\begin{figure*}
    \centering
    \includegraphics[width=\textwidth]{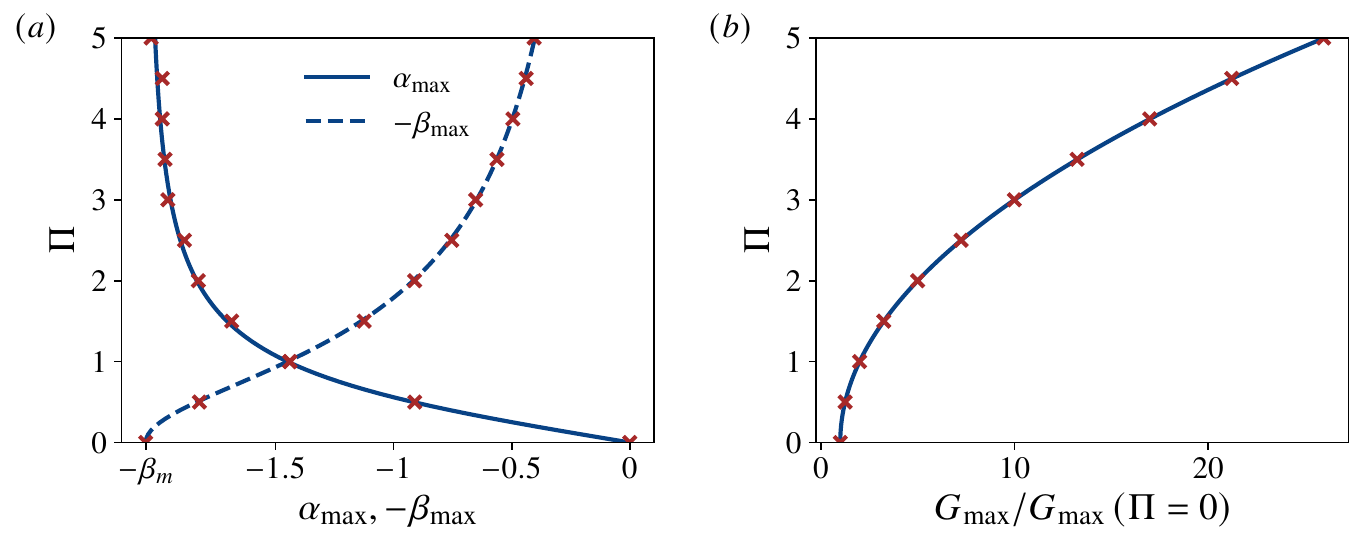}
    \caption{$(a)$, the wavenumber pair yielding $G_{\max}$ for various $\Pi$ -- note that we have chosen $\beta_m = 2.05$ here; $(b)$, the optimal amplification $G_{\max}$ normalized by its value for Poiseuille flow $\left(\Pi = 0\right)$; from Equations (\ref{eqn:gmax_ppf}) and (\ref{eqn:gmax_tsc}), this should roughly scale as $\sim 1+\Pi^2$. In either panel, the markers denote data from numerical experiments.}
    \label{fig:gmax_params}
\end{figure*}

To preface this section, we begin with an analysis analogous to the derivation of Equation (\ref{eqn:reduced_evp}). In particular, converting to our modified coordinate system $\vec{x}^*$ implies 
\begin{equation}
    G\left(\alpha, \beta, Re, U, W, t\right) = G\left(\alpha^*, \beta^*, Re, U^*, 0, t\right)
\end{equation}
With some additional algebra, one concludes
\begin{equation}
\label{eqn:transform_g}
    G\left(\alpha^*, \beta^*, Re, U^*, 0, t\right) = G\left(\alpha^*, \beta^*, Re^*, U, 0, t^*\right)
\end{equation}
where $t^* = \Pi^* t$. From here, we investigate the quantity $G_{\max}$, defined through the following optimization
\begin{equation}
    G_{\max} = \max_{\alpha,\beta, t}G\left(\alpha,\beta, Re, \Pi, t\right)
\end{equation}
In words, $G_{\max}$ represents the largest possible amplification in time and across wavenumber space -- for purely streamwise flows ($W = 0$), $G_{\max}$ is well-known to be achieved by streamwise-invariant disturbances, $\alpha = 0$. Indeed, specifically for Poiseuille flow, this maximal amplification is attained for $\left(\alpha_{m}, \beta_{m}\right) \approx \left(0, \pm 2.05\right)$, with $G_{\max}$ obeying the following asymptotic relation \cite{tref_pseudospec}
\begin{equation}
\label{eqn:gmax_ppf}
    G_{\max}^{\text{Poiseuille}}\sim \left(\dfrac{Re}{71.5}\right)^2
\end{equation}
Furthermore, the time $t_{\max}$ at which the maximal gain is realized satisfies
\begin{equation}
\label{eqn:tmax_ppf}
    t_{\max}^{\text{Poiseuille}} \sim \dfrac{Re}{13.2}
\end{equation}
Then, from Equation (\ref{eqn:transform_g}), we conclude that for TSC flows, the optimal amplification $G_{\max}$ occurs for $\left(\alpha^*_{\max}, \beta^*_{\max}\right) = \left(\alpha_{m},\beta_{m}\right)$. Of course, in the original coordinate frame, this corresponds to
\begin{equation}
    \begin{pmatrix}
        \alpha_{\max} \\ \beta_{\max}
    \end{pmatrix} = \dfrac{1}{\sqrt{1+\Pi^2}}\begin{pmatrix}
        -\Pi\beta_m \\ \phantom{-}\beta_m
    \end{pmatrix}
\end{equation}
indicating an increasing preference for the amplification of oblique disturbances as the cross-flow becomes stronger. Furthermore, from Equations (\ref{eqn:gmax_ppf}) and (\ref{eqn:tmax_ppf}), one can determine that
\begin{equation}
\label{eqn:gmax_tsc}
    G_{\max}\sim \left(\dfrac{Re^*}{71.5}\right)^2 = \left(\dfrac{Re\sqrt{1+\Pi^2}}{71.5}\right)^2
\end{equation}
and that
\begin{equation}
    t_{\max} = t_{\max}^{\text{Poiseuille}}
\end{equation}
Figure \ref{fig:gmax_params} verifies these deductions, which become more accurate as $Re$ increases.

\begin{figure*}
    \centering
    \includegraphics[width=\textwidth]{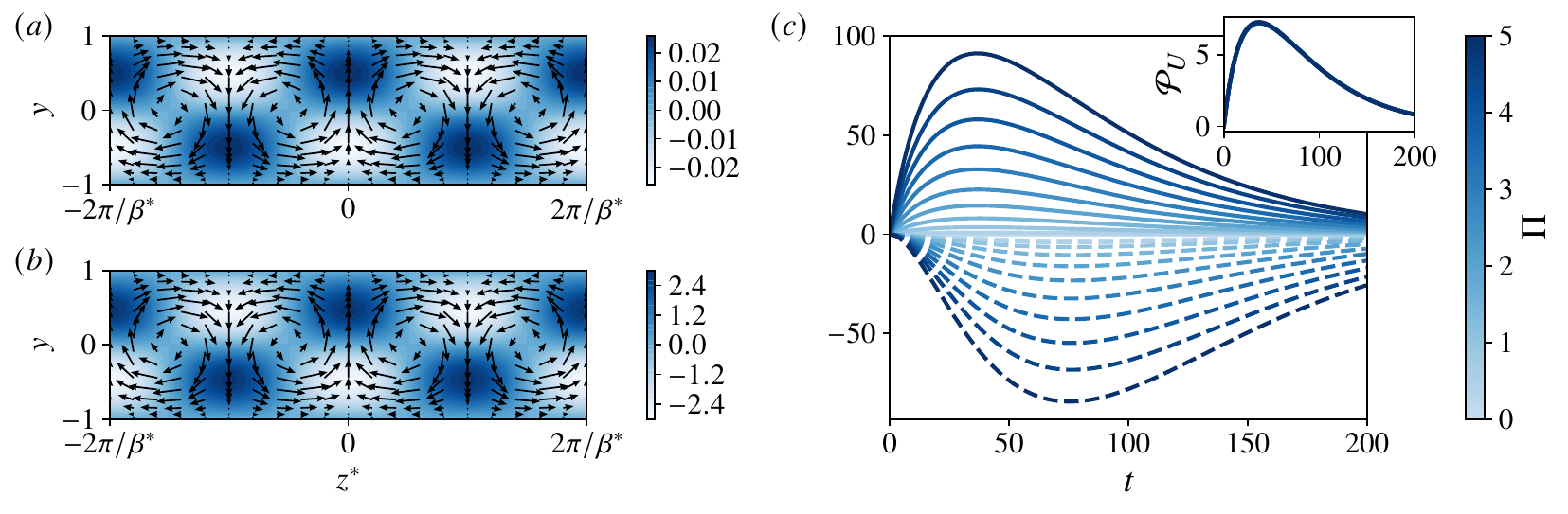}
    \caption{$(a)$ and $(b)$, the initial condition and response field (at $t = t_{\max}$) associated with $G_{\max}$ at $\Pi = 2.5$ -- color/quivers denote the streamwise/cross-stream components of the disturbance; $(c)$: for select $\Pi$, the time evolution of the linear budget for the energy-optimal initial condition. In the main panel, the solid line denotes $\mathcal{P}_W$ and the dashed line $\varepsilon$.}
    \label{fig:gmax_ic_and_budget}
\end{figure*}

The energy-optimal initial condition/response pair can be obtained as the first right and left singular vectors of the (weighted) state transition operator, respectively. For purely streamwise flows, this disturbance is well-known to develop via the lift-up effect \cite{ellingsenpalm, schmidstability}. Within the framework we have developed in this work, it is clear that a similar mechanism must operate for TSC flows, except along the modified coordinate basis $\vec{x}^*$. This is confirmed in Figures \ref{fig:gmax_ic_and_budget}$(a)$ and \ref{fig:gmax_ic_and_budget}$(b)$: streamwise-oriented streaks present at initial time intensify due to the wall-normal transport of mean horizontal momentum. In addition, one can investigate the time evolution of the total perturbation energy $E$, which evolves according to the following linearized budget
\begin{equation}
\label{eqn:energy_budget}
    \dfrac{\mathrm{d}E}{\mathrm{d}t} = \mathcal{P}_U + \mathcal{P}_W - \varepsilon
\end{equation}
where
\begin{equation}
    E = \dfrac{1}{2}\int_\mathcal{V} \left(\vec{u}\cdot\vec{u}\right)\,\mathrm{d}\mathcal{V}
\end{equation}
and $\mathcal{V}$ is the domain volume (chosen as one disturbance wavelength in accordance with our choice of the energy norm). In Equation (\ref{eqn:energy_budget}), $\mathcal{P}_U$ and $\mathcal{P}_W$ represent production against the background streamwise and spanwise shears, respectively, while $\varepsilon$ denotes the rate of viscous dissipation. Figure \ref{fig:gmax_ic_and_budget}$(c)$ plots the development of these terms in time for the energy-optimal initial condition associated with various $\Pi$. We immediately see that the streamwise production $\mathcal{P}_U$ remains essentially agnostic to the strength of the cross-flow. Moreover, the bulk of the contribution to the (significantly enhanced) energy production comes from the spanwise term $\mathcal{P}_W$, which even for moderate $\Pi$, tends to dominate by roughly an order of magnitude. The rate of viscous dissipation also increases with $\Pi$, becoming relevant, however, only at later times i.e., after significant linear growth has already occurred.

\section{\label{sec:conclusions} Conclusions \& Discussions}

To further our understanding of transition in three-dimensional boundary layers, we have investigated here for the first time the modal and non-modal stability in transversely strained channel flows. Through the introduction of a rotated coordinate basis and, subsequently, identification with the Orr-Sommerfeld-Squire system for Poiseuille flow, we have shown how the linear stability of TSC flows can be completely prescribed using previously known results.

In terms of modal growth, the critical Reynolds number is found to decrease monotonically with the strength of the cross-stream pressure gradient. On the other hand, the critical (streamwise) phase speed increases to a singularity as $\Pi\to\infty$. However, with an appropriate re-scaling of the relevant parameters in this limit, it is shown that the stability characteristics of TSC flows do in fact approach those of the Poiseuille flow, corroborating physical intuition. Insofar as unstable eigenmodes are concerned, for sufficiently large $\Pi$, the spanwise component of these perturbations appears to approach a structure similar to the usual Tollmien-Schlichting wave.

Within the framework of non-modal growth, we show first that at fixed $Re$, the optimal amplification $G_{\max}$ must increase (quadratically, in fact) with $\Pi$. This growth is nevertheless achieved in the same amount of time as a standard channel flow, indicating a greater inclination toward the onset of non-linearity and, consequently, transition. The energy-optimal initial condition develops via the lift-up process, similar to two-dimensional flows, with the only exception that this disturbance must be oriented along the direction of the net flow. The linearized energy budget is also investigated in this regard, and it is found that the bulk of the contribution to the perturbation energy is due to the production against the spanwise shear. Even for intermediate $\Pi$, the latter can rapidly increase to within multiple orders of magnitude of $\mathcal{P}_U$ (which itself remains invariant). 

\begin{acknowledgments}
All numerical work for this article was performed on the Bates College Leavitt Cluster.
\end{acknowledgments}

\bibliography{apssamp}

\end{document}